# Rearrangement of the structure during nucleation of a cordierite glass doped with TiO$_2$

Marie Guignard[1], Laurent Cormier[1], Valérie Montouillout[2,3], Nicolas Menguy[1], Dominique Massiot[2,3], Alex C. Hannon[4], Brigitte Beuneu[5]

[1] Institut de Minéralogie et de Physique des Milieux Condensés IMPMC, CNRS UMR 7590, Université Pierre et Marie Curie, Université Paris Diderot, 140 rue de Lourmel, 75015 Paris, France

[2] Conditions Extrêmes et Matériaux : Haute Température et Irradiation CEMHTI, CNRS UPR 3079, 1D avenue de la Recherche Scientifique, 45071 Orléans Cedex 2, France

[3] Université d'Orléans, Faculté des Sciences, avenue du Parc Floral, BP 6749 45067 Orléans cedex 2, France

[4] ISIS Facility, CCLRC Rutherford Appleton Laboratory, Chilton, Didcot, Oxon OX11 OQX United Kingdom

[5] Laboratoire Léon Brillouin, C.E.A. Saclay, 91191 Gif-sur-Yvette, France

E-mail: Laurent.Cormier@impmc.upmc.fr

**Abstract**

Ordering of disordered materials occurs during the activated process of nucleation that requires the formation of critical clusters that have to surmount a thermodynamic barrier. The characterization of these clusters is experimentally challenging but mandatory to improve nucleation theory. In this paper, the nucleation of a magnesium aluminosilicate glass containing the nucleating oxide TiO$_2$ is investigated using neutron scattering with Ti isotopic substitution and $^{27}$Al NMR. We identified the structural changes induced by the formation of crystals around Ti atoms and evidenced important structural reorganization of the glassy matrix.

**1. Introduction**

The appearance of an disorder-order transition is the obvious phenomenon achieved during nucleation processes [1]. Nucleation, i.e. formation of critical clusters capable of growing to form ordered structures, plays a central role in casting of metals and their alloys, folding of the proteins and crystallization of glasses [2-7], and it has attracted considerable theoretical and experimental interests [2,8,9]. This arises since the Classical Nucleation Theory (CNT) is one of the few areas of science in which several orders of magnitude are usually observed between predicted and measured rates. The mechanism by which a new stable phase appears in an initially homogeneous undercooled liquid requires a better understanding at the atomic level of fluctuations that will induce long-range oscillations in the average density. The challenge lies in the experimental difficulties to characterize the critical nuclei and in particular the composition, the density, the structural order of the nuclei compared to the bulk crystalline phase or the undercooled liquid/nuclei interface [1,10-14].

Oxide glasses provide ideal media for investigating nucleation phenomena in a condensed system and they have been widely used for quantitatively testing the applicability of standard nucleation theories [8,14]. In these materials, the studies are facilitated by slow crystal growth rates (in contrast to the freezing of liquids), which allow isolation of the early stages of transformation. The "frozen in" of the nucleation provides a unique opportunity for experimental observations.

Adding a small concentration of impurities can affect the nucleation rate dramatically. This property is widely used to develop glass-ceramic, in which special nucleating agents incorporated into the parent glass act as a catalyst for nucleation [15,16]. Due to the mixing of the glassy matrix and the crystals, glass-ceramics are advanced materials with original properties for important technical, consumer, aerospace, medical or biological applications. The nucleating agents promote bulk nucleation by accelerating phase separation or by lowering the energy barrier of nucleation. Again the underlying role of these agents is poorly understood due to the lack of atomic scale information, preventing the full identification of the catalytic sites, the structure of the critical nuclei and their potency to give the thermodynamic driving forces towards crystal growth.

In this paper, we report a detailed experimental investigation of the structural modifications arising in the first stages of nucleation of an oxide glass. The glass composition chosen for this study is $2MgO-2Al_2O_3-5SiO_2-TiO_2$ (Ti-COR) which is an archetypal glass-ceramics having important technological applications.

## 2. Experimental procedures

### 2.1. Sample preparation

Two Ti-COR glasses have been prepared by melting dried starting materials (MgO, $Al_2O_3$, $SiO_2$ and $TiO_2$) [17]. The first sample was enriched with $^{46}TiO_2$ (70.2 %) and the second one with $^{48}TiO_2$ (97.7 %). The powders were mixed and melted one hour at 1600°C in a platinum crucible. Glasses were obtained by immersing the bottom of the crucible into water. The obtained glasses were ground and melt once again to ensure a good homogeneity. The two glasses were then heat treated at ~45 °C above their glass transition temperature (Tg = 750 ± 2°C) during 4h and 12h to promote the formation of nuclei.

Eleven phases in the $MgTi_2O_5$ -$Al_2TiO_5$ (MAT) solid solution were synthesized with the molar compositions $(MgTi_2O_5)_x$-$(Al_2TiO_5)_{1-x}$ for which x is comprised between 0 and 1. Starting materials (MgO, $Al_2O_3$ and $TiO_2$) were finely ground together in ethanol and the resulting powders were pressed to make pellets. The pellets were then 24 hours at 1500°C and quenched to room temperature.

### 2.2. Neutron diffraction with isotopic substitution

The neutron scattering (NS) experiments for the two glasses were carried out at room temperature before and after the heat treatment using the GEM Diffractometer (0.1 Å$^{-1}$ < Q < 60 Å$^{-1}$) at ISIS (Rutherford Appleton Laboratory, UK) for the sample heat treated 4 hours and using the 7C2 diffractometer (0.6 Å$^{-1}$ < Q < 15.4 Å$^{-1}$) at the Laboratoire Léon Brillouin (France) for the sample heat treated 12 hours. The advantage of the isotopic substitution contrast technique is to separate specific Pair Distribution Functions (PDFs) (or sum of PDFs) yielding a chemical probe at both short and intermediate range order [18]. It was shown in a previous paper that the NS experiments with Ti isotopic substitution lead to the determination of two meaningful difference correlation functions [17]. The first correlation function that contains only the atomic PDFs involving titanium is the "Ti-centered'" correlation function $\Delta D^{Ti}(r)$. The second correlation function that is essentially free of all the atomic PDFs involving titanium is the "Ti-free'" correlation function $\Delta D^{noTi}(r)$, that contains the Si-, Al- and Mg-centered PDFs. These functions are more informative than the total correlation functions that contain overlapping contributions from all PDFs.

*2.3. Transmission Electron microscopy*

Transmission Electron Microscopy (TEM) observations were carried out on a JEOL 2100F microscope that was operating at 200 kV, equipped with a field emission gun, a high resolution UHR pole piece and a Gatan US4000 CCD camera. X-ray Energy Dispersive Spectroscopy (XEDS) analyses were used to obtain elemental mapping using a JEOL detector coupled with a scanning TEM device. The three window technique [19] allows to get Energy Filtered TEM (EFTEM) which was obtained with a GATAN Imaging filter 2001.

2.4. $^{27}$Al Nuclear Magnetic Resonance

The $^{27}$Al high-resolution NMR spectra were performed with a high field Bruker AVANCE spectrometer (17.6 T - 750 MHz) equipped with high speed MAS probeheads (spinning rate of 30kHz, aluminum free zirconia rotors of 2.5 mm diameter). The $^{27}$Al 1D spectra have been acquired using a one pulse sequence. The pulse angle was sufficiently small and the recycling delay (1s) sufficiently long to ensure quantitative interpretation. The Multiple Quantum Magic Angle Spinning (MQMAS) [20] experiments have been recorded using the shifted-echo pulse sequence with acquisition and processing of the full echo [21] and synchronized acquisition of the indirect dimension [22]. The triple quantum excitation and conversion where achieved under high power irradiation ($\nu_{rf}$ ~150 kHz) and the shifted-echo generation with low power pulse ($\nu_{rf}$ ~12 kHz).

2.5. X-ray diffraction

X-ray diffraction (XRD) patterns were recorded for both the heated treated glass samples and the crystalline powders in the MAT solid solution for a 2θ-range from 15° to 125° using a Bragg-Brentano diffractometer operating at the Cu(Kα) wavelength (Panalytical X'Pert PRO MPD, equipped with a Pixel detector). Crystallized weight fraction in the two heat treated glass sample was determined by refining their XRD pattern using the Rietveld method with internal standard [23] (Fluorite $CaF_2$ was used as the internal standard as its few Bragg peaks do not overlap with that of MAT phases). Cell parameters *a*, *b* and *c* of the orthorhombic phases in the MAT solid solution were extracted for all the XRD patterns using the Le Bail method [24]. Both the Rietveld refinement and the Le Bail fitting were done using the Fullprof suite [http://www.ill.eu/sites/fullprof/].

## 3. Results and discussion

### *3.1. Modification of the Ti environment upon crystallization*

The NS difference correlation functions are shown in figure 1 and figure 2, for the samples heat treated for 4 hours and 12 hours, respectively. They are compared to those for the Ti-COR glass before the heat treatment. The experimental data in the reciprocal space were obtained from the GEM diffractometer over a wider Q-range than those obtained from the 7C2 diffractometer. This explains that a better resolution in the real space is achieved for the correlation functions in figure 1 than for those in figure 2. Though the different resolution can modify details of the $\Delta D(r)$ functions, we are concerned in this paper of important modification of the peaks with the nucleation that are well beyond the changes due to experimental resolution.

Clearly the major changes are observed for the $\Delta D^{Ti}(r)$ functions rather than $\Delta D^{noTi}(r)$, indicating that the nucleation takes place around Ti atoms, even after a short heat treatment. This is consistent with previous studies showing that nucleation begins with the formation of a magnesium aluminotitanate ordered phase [15]. The first peak of the $\Delta D^{Ti}(r)$ functions corresponds to the first Ti-O distance. After the heat treatment, this peak is slightly modified and shifted (by ~0.02 Å) toward higher distances for the 12 hours heat treatment. The most visible structural rearrangement in the Ti environment appears above 2.5 Å. The peak at ~3.35 Å in the $\Delta D^{Ti}(r)$ functions of the parent glass results from the contributions of the Ti-M (Si, Al, Mg or Ti) atomic pairs, among those Ti-Si and Ti-Al pairs are predominant due to the large amount of silicon and aluminum atoms. After the nucleation, this peak is shifted to ~3.61 Å independently of the duration of the heat treatment. As the third peak at ~4.4 Å, it corresponds to the Ti-O second distances and it is shifted toward lower distances after the heat treatment. In the broad peaks around 6.4 Å and 9.0 Å for the parent glass, new intense and narrow contributions are emerging as the nucleation proceeds. This is a clear evidence of the structural ordering around Ti.

*3.2. Determination of the nano-crystalline phases*

Transmission electron microscope (TEM) observations confirmed the absence of nanometer-size heterogeneities (crystalline or amorphous) before the heat treatment. This was also confirmed with our multi-techniques characterization of the parent glass in a previous study [17].

To better characterize the initial nano-crystallites involving Ti we have used TEM. Dark field imaging shows that the initial nano-particles crystallize with a random distribution within the whole volume of the sample (figure 3a). We have seen no evidence of phase separation at early stages of heat treatment, in agreement with a recent high resolution TEM investigation [25]. High Resolution TEM observations obtained from these particles (figure 3b) confirm that the first phases that crystallize in the parent glass belong to the $MgTi_2O_5$-$Al_2TiO_5$ solid solution phase [15]. Due to their sizes that do not exceed 20-30 nanometers and the presence of the surrounding glassy matrix, the exact composition of these crystals cannot be obtained. X-ray diffraction (XRD) and Rietveld analysis was therefore used to determine the chemical composition of the nano-crystals and the weight fraction of the crystalline phase.

Lattice parameters $a$, $b$ and $c$ for the orthorhombic phases in the $MgTi_2O_5$-$Al_2TiO_5$ solid solution are plotted in figure 4 as a function of the molar percentage of $MgTi_2O_5$. They vary linearly with composition, closely following Vegard's law which applies to solid solutions formed by random substitution ions. The lattice parameters for the nano-crystals that precipitated in the Ti-COR glass during the heat treatment were determined from the Rietveld refinement giving $a = 3.647 \pm 0.005$ Å, $b = 9.607 \pm 0.008$ Å and $c = 9.885 \pm 0.008$ Å. When reported on figure 4, these cell parameters do not lead to a unique chemical composition for the nano-crystals. The parameter $a$ is compatible with a molar composition close to $(MgTi_2O_5)_{40}$-$(Al_2TiO_5)_{60}$, while the parameters $b$ and $c$ are compatible with the compositions $(MgTi_2O_5)_{55}$-$(Al_2TiO_5)_{45}$ and $(MgTi_2O_5)_{68}$-$(Al_2TiO_5)_{32}$ respectively.

This can be explained by the fact that these nano-particles were obtained by heat treatment of the Ti-COR glass at 795°C and frozen in by quenching to room temperature. Their relative volume variation in the temperature-range 795-20°C is expected to be close to that for $MgTi_2O_5$ phase, that is around $-2.1 \times 10^{-2}$ [26-28], and that of $Al_2TiO_5$ phase, that is around -2.2×10$^{-2}$ [26,29]. However the nano-crystals represent only a few weight fraction of the heat treated glass samples, 1.4% and 4.2% for the 4 hours- and the 12 hours heat treatments respectively (The weight fraction of the crystalline phase in the heat treated samples is reported in Table 1). Therefore the relative volume variation of the nano-particles was imposed by the volume variation of the remaining glass after the heat treatment in the temperature-range 795-20°C which is approximately -1.0×10$^{-2}$ for a thermal expansion coefficient of ~4.5×10$^{-6}$ K$^{-1}$ [30]. This implies that the nano-crystals are subject to tensile stress and therefore their lattice parameters determined at room temperature differ to the bulk ones.

The thermal expansion coefficient for the parameter *a* is 2.3×10$^{-6}$ K$^{-1}$ for the MgTi$_2$O$_5$ phase [26], while it is around -3×10$^{-6}$ K$^{-1}$ for the Al$_2$TiO$_5$ phase [26,29]. Therefore it should be null or very small for most of the compositions in the MAT solid solution. As the lowest expansion is always found in the *a*-direction in MAT crystals [26], we supposed that the parameter *a* for the nano-crystals embedded within the Ti-COR glass-ceramic is the same as that of the crystalline powders. Therefore the molar composition of nano-crystals was assumed to be close to (MgTi$_2$O$_5$)$_{40}$-(Al$_2$TiO$_5$)$_{60}$. The XRD pattern for the crystalline powder with this chemical composition was recorded *in situ* at 795°C and their lattice parameters were determined from Le Bail fitting [24] of these experimental data. We obtained *a* = 3.6469 ± 0.0002 Å, *b* = 9.6479 ± 0.0003 Å and *c* = 9.9300 ± 0.0003 Å. Firstly, this confirms that the thermal expansion coefficient for the parameter *a* is null for the (MgTi$_2$O$_5$)$_{40}$-(Al$_2$TiO$_5$)$_{60}$ phase in the temperature-range 20-795°C. Secondly, the lattice parameters at 50°C and 795°C for (MgTi$_2$O$_5$)$_{40}$-(Al$_2$TiO$_5$)$_{60}$ gives a relative volume variation of -0.9×10$^{-2}$, which is compatible with that of the remaining glass matrix. We can thus give a determination of the first nano-crystals appearing during the heat treatment with the formation of the pseudobrookite-type phase (MgTi$_2$O$_5$)$_{40}$-(Al$_2$TiO$_5$)$_{60}$.

### 3.3. Role of the high coordinated Al atoms

As a next step, we have performed $^{27}$Al high-resolution NMR experiments to quantify the distribution of the coordination number for aluminum atoms (Details of the experiments can be found in Ref. [17]). The $^{27}$Al NMR 1D Magic-angle Spinning (MAS) spectra and the 2D Multiple Quanta Magic-angle Spinning (MQ-MAS) spectra are presented in figure 5. They were deconvoluted using the DMfit software [31] and the coordination numbers are reported in Table 1. In alkaline-earth alumino-silicate glasses, aluminum atoms usually substitute for silicon atoms in tetrahedral sites when they are associated to alkaline earth ions $M^{2+}$ that ensure the charge balance of the negative charge of the $(AlO_4)^-$ tetrahedra. In the Ti-COR glass, fivefold- ($^{[5]}Al$) and sixfold-coordinated ($^{[6]}Al$) aluminum atoms are also present (Table 1). The existence of these highly coordinated aluminum atoms has been already reported in MgO-Al$_2$O$_3$-SiO$_2$ glasses in which Mg$^{2+}$ ions may not be fully available to compensate the $(AlO_4)^-$ tetrahedral [32,33].

The intensity of the peak related to $^{[6]}Al$ atoms gradually increases in the $^{27}Al$ 1D NMR spectra (figure 5) as the duration of the heat treatment increases. The amount of $^{[6]}Al$ atoms is partially correlated with the formation of MAT nanocrystals. However the proportion of $^{[6]}Al$ atoms determined by $^{27}Al$ NMR analysis is much higher than the proportion of aluminum atoms localized in the octahedral sites of the MAT nanoparticles determined using XRD (Table 1). This indicates that the nucleation in the Ti-COR glass does not only imply the formation of the nanocrystals but it also modifies the local order in the amorphous part of the sample (or at the glass/nuclei interface). Moreover, the $^{[5]}Al/^{[4]}Al$ ratio determined by NMR analysis decreases as the duration of the heat treatment increases. In a previous investigation, we have shown that $^{[5]}Al$ and Ti atoms were preferentially linked by edge-sharing, a structural arrangement that mimics the MAT phase and could then be seeds for nucleation [17]. Therefore the decrease of the $^{[5]}Al/^{[4]}Al$ ratio shows that $^{[5]}Al$ atoms are preferentially involved in the formation of the nuclei and that they likely play an important role during the nucleation process.

NS experiments also confirm that local order is modified after the heat treatment. A slight widening of the first peak in the $\Delta D^{noTi}(r)$ functions is observed around 1.8-1.9 Å (figures 1b and 2b). This can be attributed to the increase of the Al-O bond length correlated with the increase of the amount of $^{[6]}Al$ during the heat treatment because $^{[6]}Al$-O bonds are generally longer than $^{[4]}Al$-O or $^{[5]}Al$-O ones (for example in alumino-silicate crystals, $d_{[4]Al-O} \sim 1.74$ Å and $d_{[6]Al-O} \sim 1.91$ Å) [34,35].

### 3.4. Modification of the glass structure with the formation of the MAT nano-crystals

In the MAT crystals, Mg, Al and Ti atoms occupy the two non-equivalent octahedral sites with a partial cation order [36]. The chemical composition of the MAT phase implies a Al/Mg ratio of 3 in the nano-crystals compared to 2 in the parent glass. Therefore, the crystalline particles are slightly enriched in aluminium with respect to the glass matrix.

Though the nucleation only involved a MAT phase, the $\Delta D^{noTi}(r)$ function shows substantial modifications above 3.5 Å after the nucleation, simultaneously with the transformation around Ti. The main effect of the heat treatment is the appearance of a new ordering in the intermediate range order, with (1) an increase in intensity and a shift toward lower distances of the peak at ~5.1 Å, that corresponds to O-O second distances, and (2) the splitting of contributions comprised between 6-8 Å into two distinct peaks at 6.4 Å and 7.4 Å. This is the first evidence that the initial stages of nucleation also imply profound structural changes for the remaining glassy network.

The correlation functions due to the MAT nanocrystal were numerically calculated using the Debye scattering formula on nanoparticles of finite sizes (~1 nm) [37,38]:

$$I(Q) = \sum_i \sum_j f_i f_j \frac{\sin(Qr_{ij})}{Qr_{ij}} \qquad (1)$$

where $f_i$ is the atomic scattering factor of the atom i, $r_{ij}$ is the distance between the atoms *i* and j, and the momentum transfer $Q = 4\pi \sin\theta/\lambda$ depends of the scattering angle θ and the wavelength of the incident X-rays, λ.

The simulation (plain lower curve, figure 2a) shows a qualitative good agreement with the peaks that emerge or shift upon nucleation in the $\Delta D^{Ti}(r)$ function. Therefore the appearance of these peaks after the heat treatment can directly be correlated with the presence of the MAT nanocrystals dispersed in the glass matrix. The analysis of the different PDFs indicates that the main contribution comes from the Ti-O pair and that Ti-Ti, Ti-Al and Ti-Mg pairs make weak contributions between 2.7-3.5 and 5-5.5 Å. The first mean Ti-O distance appears at 1.89 Å in the parent glass and increases to 1.91 Å in the 12 h heat treated glasses, but remains lower than in the MAT nanocrystals, $d_{[6]Ti-O}$ = 1.93 Å (figure 2a). This indicates that, in the heat treated glass, not all the Ti atoms are in six-fold coordination sites or the titanium environment is distorted compared to that expected in MAT crystals. Similarly, the discrepancies in peak positions or intensities signify that not all Ti atoms participate to the MAT nanocrystals or that these nanocrystals present a distorted structure compared to the bulk one, possibly due to strain effects on the nanoparticles embedded in the glassy matrix [39]. This new experimental approach is thus promising to better characterize the first nucleating phases and ordering occurring in the course of nucleation.

A simulation of the $\Delta D^{noTi}(r)$ function with the Debye formula shows that the PDFs not involving Ti in the MAT nanocrystals (plain lower curve, figure 2b) do not allow an explanation of the structural modifications appearing upon heating. The peaks that increase in intensity at 4.95, 6.4 and 7.4 Å in the experimental function does not coincide with specific distances expected in the nanocrystals. Despite the small crystalline weight percent (1-4%, Table 1), our results clearly show that nucleation induces strong structural changes in the remaining glassy network, an effect that has never been observed yet.

A nucleation mechanism emerges from these results. The first nano-crystals have a composition close to $(MgTi_2O_5)_{40}$-$(Al_2TiO_5)_{60}$. Ti does not favor nucleation due to an inhomogeneous distribution but acts as nucleating agent due to its specific association with high coordinated Al species. Ti and Al polyhedra form edge-sharing complexes that mimic the crystalline MAT organization. Therefore, few local atomic rearrangements above the glass transition temperature are required to promote the transition from the amorphous to the crystal. These nanocrystals differ from the bulk crystals and their formation strongly affects the remaining glass structure, inducing organization in the glass in the early nucleation stages, which can lead after additional heat treatment to the formation of cordierite crystals [15].

## 4. Conclusions

We have revealed the structural rearrangements occurring in the Ti-containing cordierite glass upon nucleation. Neutron scattering data coupled with Ti isotopic substitution have shown important structural modifications in the Ti environment and difference functions agree with the initial formation of magnesium aluminotitanate nuclei, possibly due to preferential linkages between Ti and $^{[5]}$Al [17]. Simultaneously to the structural changes around Ti, we have also evidenced strong modifications in the remaining glass network, which imply an increasing order in the glass structure early in the nucleation process. This reflects the atomic rearrangements in the glassy part resulting from the diffusion of the chemical species towards the glass/crystal interface. One of these rearrangements is the modification of the Al coordination that varies from tetrahedral to octahedral as the nucleation proceeds. The resulting $^{[6]}$Al can then form new sites favoring the formation or growth of MAT nuclei. Our study emphasizes that the development of structural investigations can give new information at the atomic scale that are essential to improve the picture of nucleation processes in amorphous materials.

**Acknowledgments**

We thank Olivier Dargaud for TEM imaging and Georges Calas for helpful discussion. This works was supported by the French national research agency (ANR) under contract N°06-JCJC-0010 "Nuclevitro".

**Table 1.** $t$ represents the duration of the heat treatment (in hours). $W_\alpha$ is crystallized weight fraction obtained from the X-ray diffraction analysis. $Al_\alpha$ is the percentage of the Al atoms in the nano-crystals calculated from their chemical composition and from the crystallized weight fraction (in % of the total Al atoms in the sample). $^{[n]}Al$ is percentage of Al atoms coordinated by $n$ oxygen atoms obtained from the $^{27}Al$ NMR analysis (in % of the total Al atoms in the sample).

| t | $W_\alpha$ | $Al_\alpha$ | $^{[4]}Al$ | $^{[5]}Al$ | $^{[6]}Al$ |
|---|---|---|---|---|---|
| 0 | 0 | 0 | 83.3 ± | 14.6 ± | 2.1 ± |
| 4 | 1.2 ± 0.6 | 1.2 ± 0.7 | 80.7 ± | 13.6 ± | 5.7 ± |
| 12 | 4.1 ± 2.0 | 4.3 ± 2.3 | 72.5 ± | 5.3 ± | 22.2 ± |

Figure Caption

**Figure 1.** Neutron scattering experiments carried out on the Ti-COR glass before and after the heat treatment (4 h at 795 °C) (GEM diffractometer). (a) "Ti-centered" correlation function $\Delta D^{Ti}(r)$ (Fourier transform interval 0.6-28 Å$^{-1}$). (b) "Ti-free" correlation function $\Delta D^{noTi}(r)$ (Fourier transform interval 0.6-40 Å$^{-1}$). The first peak ~1.63 Å in the $\Delta D^{noTi}(r)$ functions corresponds to the overlap of the Si-O and Al-O distances. The shoulder observed at ~2.0 Å corresponds to the Mg-O distances.

**Figure 2.** Neutron scattering experiments carried out on the Ti-COR glass before and after the heat treatment (12 h at 795 °C) (7C2 diffractometer). (a) "Ti-centered" correlation function $\Delta D^{Ti}(r)$ (Fourier transform interval 0.6-15.4 Å$^{-1}$). The lower plain curve is the sum of the Ti-centered pair distribution functions simulated with the Debye formula for a MAT nanocrystal of ~1nm diameter. (b) "Ti-free" correlation function $\Delta D^{noTi}(r)$ (Fourier transform interval 0.6-15.4 Å$^{-1}$). The lower plain curve is the sum of the pair distribution functions not involving Ti simulated with the Debye formula.

**Figure 3.** TEM observations of the Ti-COR glass heat treated 4 h at 795 °C. (a) Overall dark field image revealing the presence of nanocrystals dispersed in the glassy matrix. (b) HREM image of a nano-crystal enclosed within the heat treated glass. The corresponding FFT (insert) is compatible with a pseudo-brookite structure observed along the [100] direction.

**Figure 4.** Lattice parameters *a* (full triangle), *b* (full square) and *c* (full circle) for the orthorhombic phases $(MgTi_2O_5)_x$-$(Al_2TiO_5)_{100-x}$ where x is the molar percentage of $MgTi_2O_5$. Dashed lines represent the linear fit for each series of data.

**Figure 5.** $^{27}$Al 1D NMR experimental spectra and the fitting models obtained with coordination numbers given in Table 1 (top) and contour plots of $^{27}$Al MQ-MAS NMR spectra (bottom) for the Ti-COR glass (a) before the heat treatment, (b) heat treated 4 h at 795 °C and (c) heat treated 12 h at 795 °C.

Figure 1

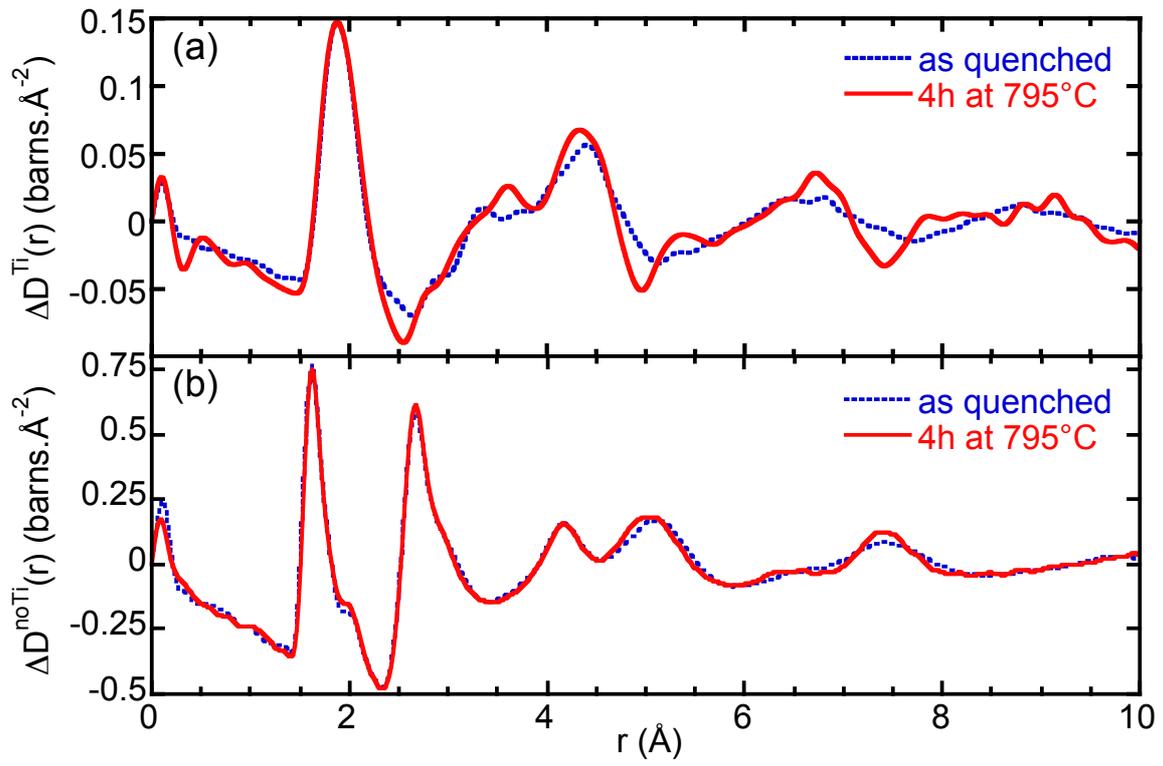

Figure 2

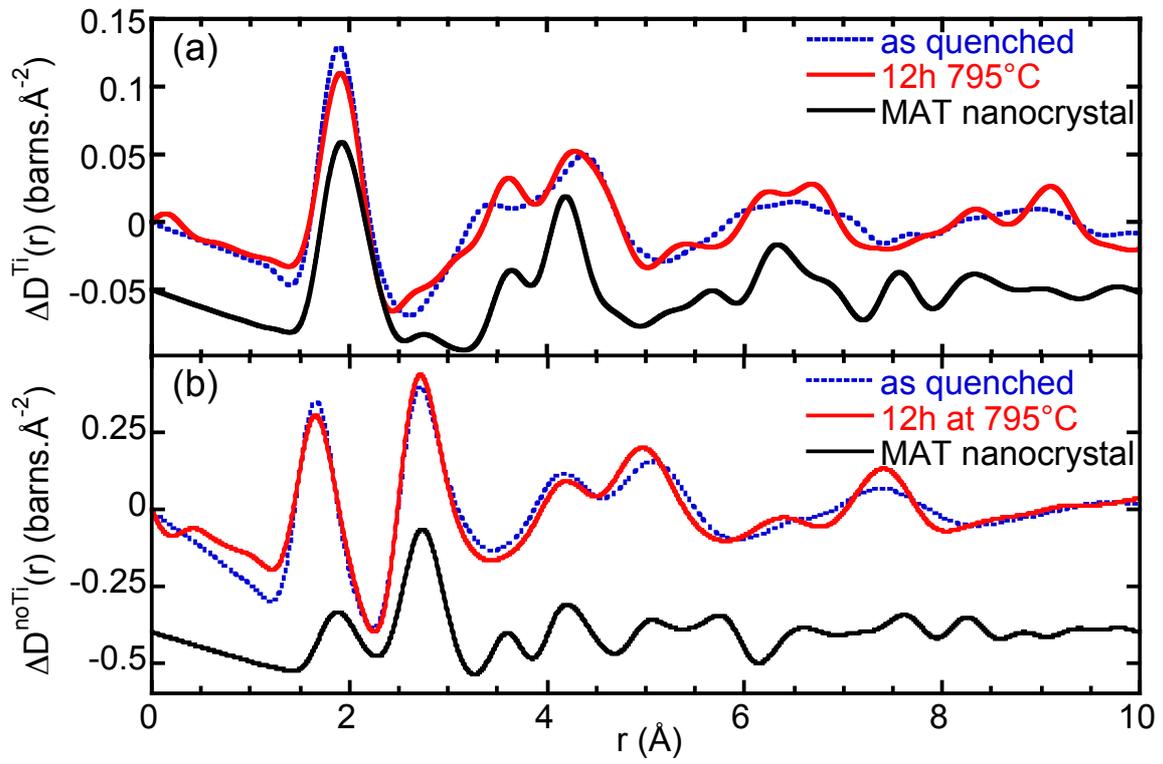

Figure 3

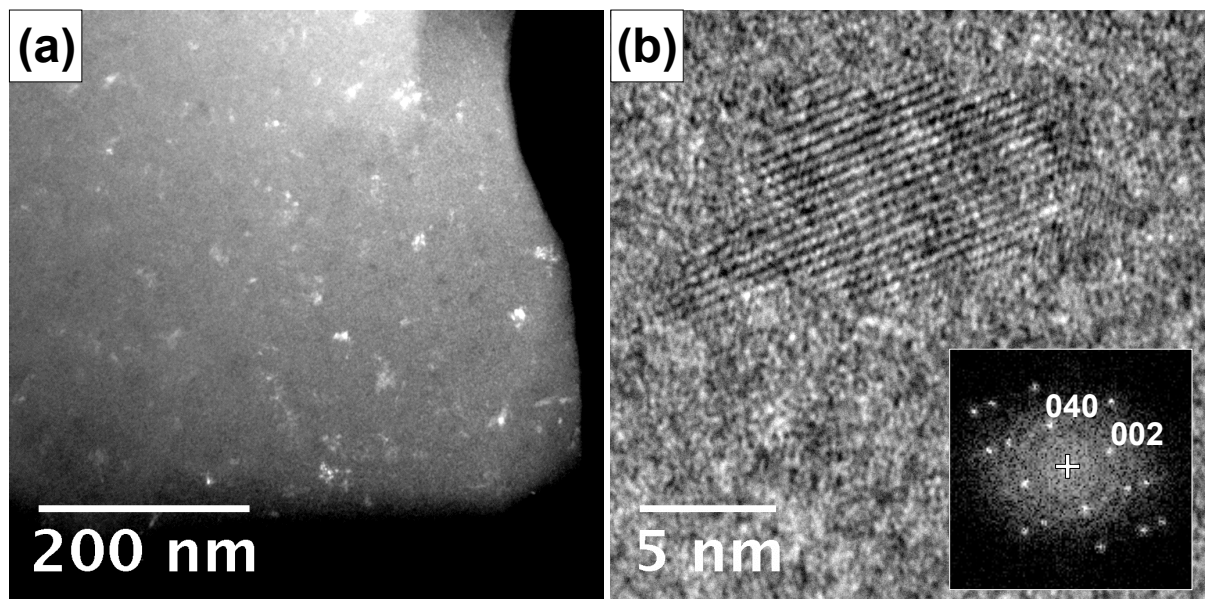

Figure 4

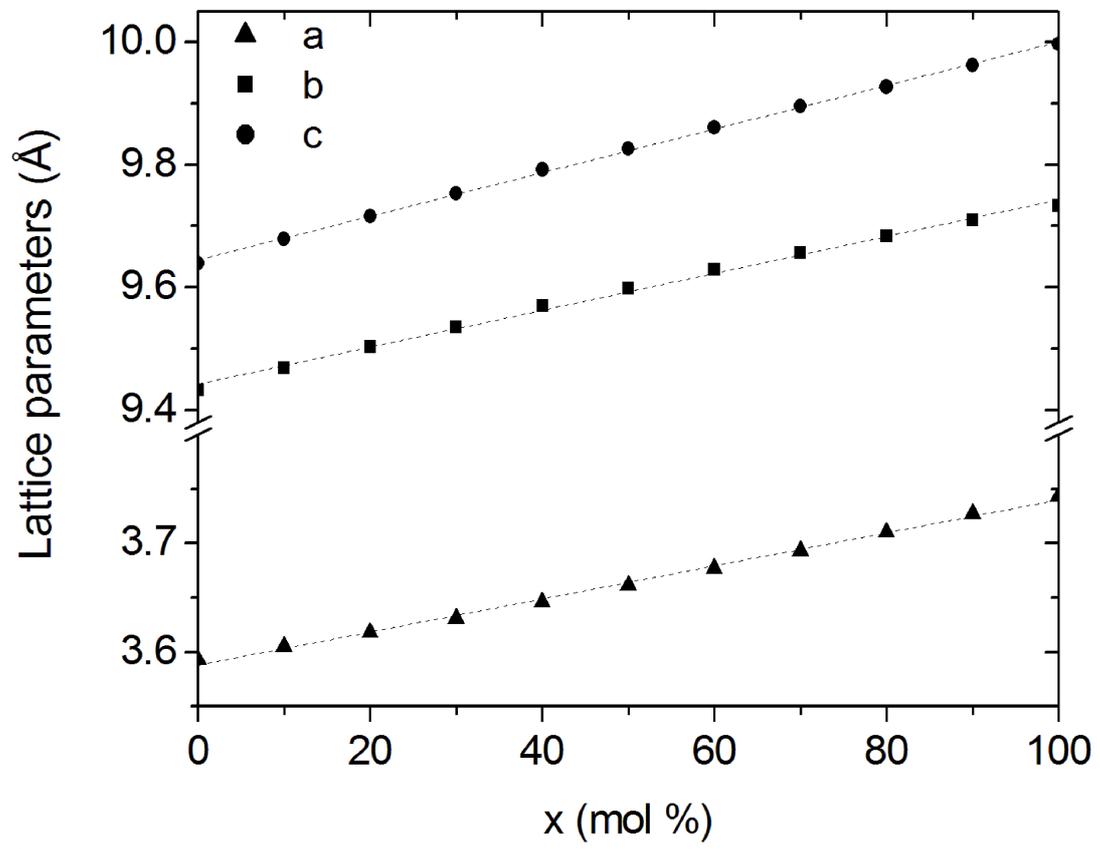

Figure 5

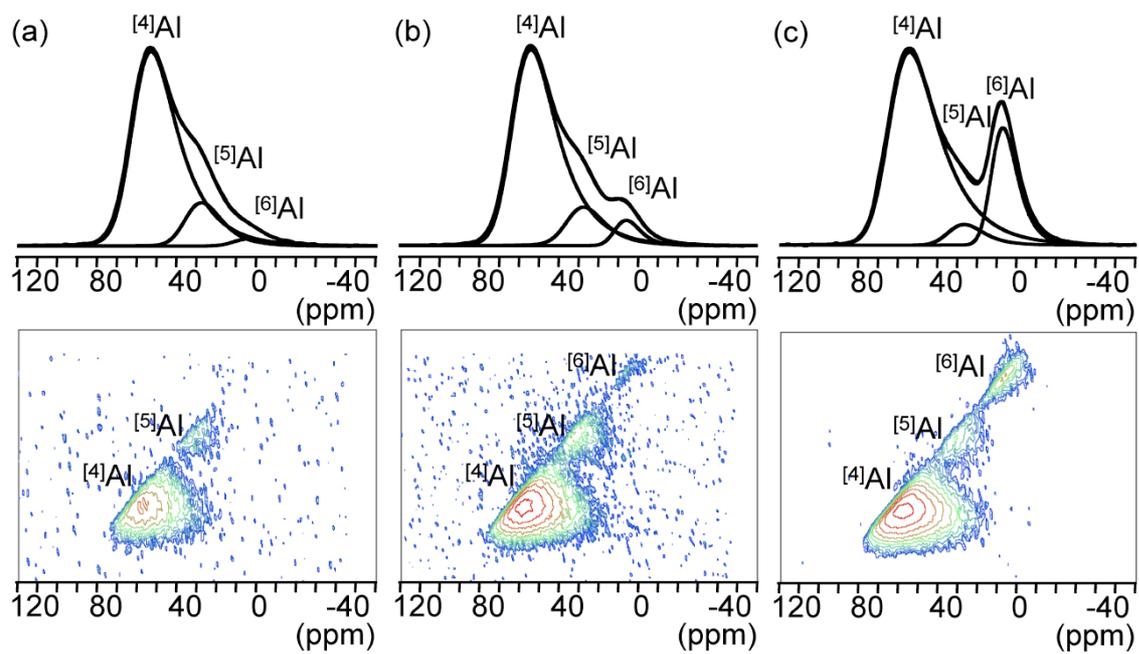